\documentclass[conference]{IEEEtran}
\IEEEoverridecommandlockouts
\usepackage{cite}
\usepackage{amsmath,amssymb,amsfonts}
\usepackage{algorithmic}
\usepackage{graphicx}
\usepackage{textcomp}
\usepackage{xcolor}
\usepackage{multirow}
\usepackage{verbatim}
\usepackage{booktabs}

\usepackage{tikz}
\usetikzlibrary{shapes, arrows}%
\tikzset{>=latex}

\usepackage{pgfplots}
\usepgfplotslibrary{fillbetween}
\usepgfplotslibrary{colormaps}
\pgfplotsset{
            compat=1.16,
            label style={font=\footnotesize},
            tick label style={font=\scriptsize},
            legend style={font=\scriptsize}
            }

\def\BibTeX{{\rm B\kern-.05em{\sc i\kern-.025em b}\kern-.08em
    T\kern-.1667em\lower.7ex\hbox{E}\kern-.125emX}}

\usepackage{tikz}
\usepackage{textcomp}
\usepackage[doipre={DOI:~}]{uri}
\usepackage{lipsum}
\newcommand\copyrighttext{%
    \footnotesize \textcopyright 2024 IEEE. Personal use of this material is permitted.  Permission from IEEE must be obtained for all other uses, in any current or future media, including reprinting/republishing this material for advertising or promotional purposes, creating new collective works, for resale or redistribution to servers or lists, or reuse of any copyrighted component of this work in other works.
    
    Accepted as a conference paper at the 2024 Biomedical Circuits and Systems Conference.
    }
\newcommand{\copyrightnotice}{%
\begin{tikzpicture}[remember picture,overlay]
\node[anchor=south,yshift=10pt] at (current page.south) {\fbox{\parbox{\dimexpr\textwidth-\fboxsep-\fboxrule\relax}{\copyrighttext}}};
\end{tikzpicture}%
}
\begin{document}
\bstctlcite{IEEEexample:BSTcontrol}

\title{Optimization and Deployment of Deep Neural Networks for PPG-based Blood Pressure Estimation Targeting Low-power Wearables}

\author{\IEEEauthorblockN{
        Alessio Burrello\IEEEauthorrefmark{1}\IEEEauthorrefmark{3},
        Francesco Carlucci\IEEEauthorrefmark{1},
        Giovanni Pollo\IEEEauthorrefmark{1},
        Xiaying Wang\IEEEauthorrefmark{2}, Massimo Poncino\IEEEauthorrefmark{1}, Enrico Macii\IEEEauthorrefmark{1},\\ Luca Benini\IEEEauthorrefmark{2}\IEEEauthorrefmark{3}, Daniele Jahier Pagliari\IEEEauthorrefmark{1}
    }
    \IEEEauthorblockA{\\\IEEEauthorrefmark{1}Dept. DAUIN, Politecnico of Turin, Italy\hspace{3.2mm}\IEEEauthorrefmark{2}Dept. ITET, ETH Zurich, Switzerland \hspace{3.2mm}\IEEEauthorrefmark{3}DEI, University of Bologna, Italy\vspace{-0.2cm}
    }

    \thanks{Corresponding email: alessio.burrello@polito.it}

    \vspace{-0.5cm}
}

\maketitle
\copyrightnotice
\begin{abstract}
PPG-based Blood Pressure (BP) estimation is a challenging biosignal processing task for low-power devices such as wearables. State-of-the-art Deep Neural Networks (DNNs) trained for this task implement either a PPG-to-BP signal-to-signal reconstruction or a scalar BP value regression and have been shown to outperform classic methods on the largest and most complex public datasets. However, these models often require excessive parameter storage or computational effort for wearable deployment, exceeding the available memory or incurring too high latency and energy consumption. In this work, we describe a fully-automated DNN design pipeline, encompassing HW-aware Neural Architecture Search (NAS) and Quantization, thanks to which we derive accurate yet lightweight models, that can be deployed on an ultra-low-power multicore System-on-Chip (SoC), GAP8. Starting from both regression and signal-to-signal state-of-the-art models on four public datasets, we obtain optimized versions that achieve up to 4.99\% lower error or 73.36\% lower size at iso-error. Noteworthy, while the most accurate SoA network on the largest dataset can not fit the GAP8 memory, all our optimized models can; our most accurate DNN consumes as little as 0.37 mJ while reaching the lowest MAE of 8.08 on Diastolic BP estimation.
\end{abstract}

\begin{IEEEkeywords}
PPG, Neural Architecture Search, Blood Pressure, DNN
\end{IEEEkeywords}

\vspace{-0.4cm}
\section{Introduction and Related Works}
\vspace{-0.2cm}
Blood pressure (BP) is a crucial health parameter that necessitates continuous monitoring for a large population of vulnerable individuals, being linked to various heart-related diseases, such as hypertension, cardiomyopathy, and heart failure~\cite{bp-disease}. Various monitoring solutions exist, from cuffless to invasive procedures, but wearable technologies such as smartwatches would enable non-invasive monitoring of larger coorts of individuals at an affordable cost and without affecting their normal lives, thereby contributing to saving many lives.. In this domain, one of the most common monitoring techniques relies on Photoplethysmography (PPG). 

PPG uses a light-emitting diode (LED) to illuminate the skin. A photodiode then collects the reflected light, whose intensity depends on the blood volume variation due to heart activity~\cite{ppg-signal}. While various medically relevant parameters can be derived from PPG, including heart rate (HR) and respiratory rate, this paper focuses on its usage for the estimation of Systolic Blood Pressure (SBP) and Diastolic Blood Pressure (DBP), reflecting the blood pressure during and in between heart muscle contractions, respectively.

A broad set of machine learning techniques have been employed in the literature for this task, ranging from classical methods like Random Forest (RF)~\cite{ppg-rf}, and Support Vector Regression (SVR)~\cite{ppg-svr} to Deep Neural Networks (DNNs)~\cite{ppg-dnn-1, ppg-dnn-2}. Further, ensemble learning frameworks have been proposed to reduce the risk of overfitting~\cite{ppg-ensemble, ppg-svr}.
In comparison to classic methods, DNNs offer the advantage of not needing an often expensive feature extraction process and have been shown to generalize better on unseen data in several biosignal processing tasks~\cite{q-ppg, burrello2022bioformers, ppg-tcn}. Several DNN architectures have been considered for PPG-based BP estimation~\cite{ppg-dnn-1,ppg-dnn-2,ppg-mlp,ppg-lstm},
with most recent works focusing on 1D Convolutional Neural Networks (CNNs)~\cite{ppg-tcn}.
Some works train these networks as regressors to directly predict a scalar DBP or SBP value based on a time window of PPG readings. ResNet-like networks achieve state-of-the-art performance in this category~\cite{resnet-ppg}. Others adopt a signal-to-signal (sig2sig) approach, where the CNN is tasked to reconstruct the entire DPB/SBP time series starting from the PPG one. In this group, architectures based on UNet~\cite{unet} are the best-performing ones. 

However, existing deep learning models for BP estimation have a large number of parameters and high computational complexity. When pursuing continuous monitoring on resource-constrained, low-power devices such as wearables, those models either exceed the available memory or incur excessive latency and energy consumption. 

This paper attempts to mitigate this issue through the use of a fully automated DNN design pipeline, which encompasses two main optimization steps, i.e., Neural Architecture Search (NAS)~\cite{nas-survey} and Quantization~\cite{jacobs2018}. 
Starting from state-of-the-art regression and sig2sig CNNs, we first apply a gradient-based NAS to automatically select each layer's operation from a pool, and tune the network depth, discovering architectures that balance BP prediction error and model size. 
We then select a subset of the Pareto-optimal DNNs identified by the NAS and quantize them to \texttt{int8} precision to further reduce their size, latency, and energy consumption. Lastly, we employ a DNN compiler~\cite{dory} to automatically convert the quantized models to optimized C code, targeting GreenWaves' GAP8 \cite{gap8}, an ultra-low-power System-on-Chip (SoC) suitable to be embedded in a wearable device for practical, efficient and continuous BP monitoring.
To the best of our knowledge, ours is the first work to apply a cost-aware NAS for BP estimation DNNs and to consider their deployment on wearable-class devices.

With experiments on four publicly available datasets~\cite{BCG_dataset,PPGB_dataset, sensors_dataset,uci_dataset}, we obtain models that reduce the BP estimation error by up to 4.99\% with respect to the best state-of-the-art DNN, or maintain the same error with up to 73.36\% fewer parameters. Furthermore, on the most complex dataset, UCI~\cite{uci_dataset}, we also outperform classic ML methods, obtaining a Mean Absolute Error (MAE) of 7.86 (versus 8.07), while also using fewer parameters.
When deployed on GAP8, our models require 7.12-8.91ms per inference while consuming 0.36-0.45 mJ. 

\vspace{-0.1cm}
\section{Materials \& Methods}

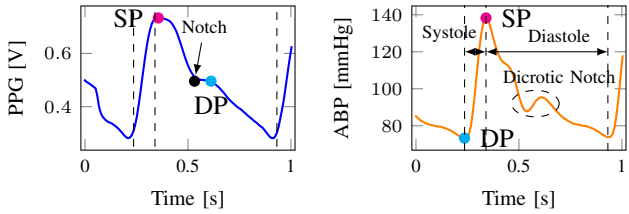
\begin{figure}[!t]
  \centering
  \begin{tikzpicture}
    \begin{axis}
    [
      no markers,
      width=0.5\columnwidth,
      height=3.5cm,
      xlabel={Time [s]},
      ylabel={PPG [V]},
      scaled ticks=false,
      xtick={62.5, 125, 187.5},
      xticklabels={},
      xticklabels={0, 0.5, 1},
      xmin = 60, xmax = 190,
    ]
      \addplot+ [name path=f, thick] table [x=x, y=ppg, col sep=comma] {images/ppg_abp_small.csv};

      \node[label={180:{SP}},circle,fill=magenta,inner sep=1.5pt] at (axis cs:107,0.732918787) {};
      \node[label={270:{DP}},circle,fill=cyan,inner sep=1.5pt] at (axis cs:139,0.495737954) {};
      \node[circle,fill,inner sep=1.5pt] (dnotch) at (axis cs:129,0.495737954) {};
      \node (snotch) at (axis cs:135,0.7){\scriptsize Notch};
      \draw[->](snotch)--(dnotch);

      \draw [dashed] (92,-1) -- (92,1);
      \draw [dashed] (105,-1) -- (105,1);
      \draw [dashed] (179,-1) -- (179,1);

    \end{axis}
    \begin{axis}
    [
    xshift=4.4cm,
      no markers,
      width=0.5\columnwidth,
      height=3.5cm,
      xlabel={Time [s]},
      ylabel={ABP [mmHg]},
      scaled ticks=false,
      xtick={62.5, 125, 187.5},
      xticklabels={0, 0.5, 1},
      xmin = 60, xmax = 190,
    ]
      \addplot+ [name path=f, thick, orange] table [x=x, y=abp, col sep=comma] {images/ppg_abp_small.csv};

      \node[label={0:{DP}},circle,fill=cyan,inner sep=1.5pt] at (axis cs:92,73.33772708) {};
      \node[label={0:{SP}},circle,fill=magenta,inner sep=1.5pt] at (axis cs:105,138.2660615) {};

      \draw [dashed] (92,-2) -- (92,150);
      \draw [dashed] (105,-2) -- (105,150);
      \draw [dashed] (179,-2) -- (179,150); %

      \draw[<->] (axis cs:92,120) -- (axis cs:105, 120) node [pos=-0.7, anchor=south] {\scriptsize Systole};
      \draw[<->] (axis cs:105,120) -- (axis cs:179, 120) node [pos=0.6, anchor=south] {\scriptsize Diastole};

      \draw[dashed] (135,92) ellipse (9pt and 5pt);
      \node[anchor=south,xshift=9pt,yshift=4pt] () at (axis cs: 135,92){\scriptsize Dicrotic Notch};
      
    \end{axis}
  \end{tikzpicture}
  \vspace{-0.35cm}
  \caption{SBP and DBP estimation from PPG (left) and ABP (right) signals.}
  \label{fig:methods:BP_PPG}
  \vspace{-0.5cm}
\end{figure}

\subsection{Blood Pressure Estimation using PPG Signal}

Blood pressure monitoring techniques can be continuous or intermittent and invasive or non-invasive. The invasive monitoring, usually performed through an intra-arterial catheter \cite{casabianca2009cardiovascular}, directly measures the atrial blood pressure signal (ABP). Common cuff-based methods like sphygmomanometer, although being gold standard and minimally invasive, are cumbersome and don't allow continuous monitoring.

On the other hand, PPG optical signal is strongly related to changes in blood volumes, and its effectiveness is already proven in various clinical applications. Although PPG is morphologically similar to ABP, and various studies have shown how the two signals share most of the informative features~\cite{correlation_study}, extracting a blood pressure estimation from it is not a trivial task, given that the signal is subject to artifacts related to movements or air between the sensor and the skin. 
Fig.~\ref{fig:methods:BP_PPG} shows examples of clean PPG and ABP signals with the points corresponding to systolic and diastolic blood pressure marked on the two plots.

\subsection{Datasets}
In our study, we adopt the same four datasets, data preprocessing, and training protocols as the extensive survey in~\cite{nature-survey}, which is state-of-the-art for this task.
All datasets are resampled to 125 Hz.
PPGBP \cite{PPGB_dataset} is the smallest dataset, with 619 total PPG segments, each lasting 2.1s, but it involves a large number of patients (218) with different cardiovascular diseases.
BCG\cite{BCG_dataset} is a bed-based ballistocardiography dataset comprising around 4 hours of cumulative measurements on 40 individuals, split into 5s windows. 
Sensors~\cite{sensors_dataset} is a subset of the MIMIC III dataset, comprising 11102 non-overlapping 5s data segments from 1195 patients.
Lastly, UCI~\cite{uci_dataset} is a subset of the MIMIC II waveform.
It's considerably bigger than all the others, with $\approx$411k segments from an unknown number of subjects.
All datasets include only measurements on resting patients in a clinical setting. Therefore, motion artifact removal using acceleration data~\cite{q-ppg,ppg-nas} can be neglected. Along with PPG signals, BCG, UCI, and Sensors provide the complete blood pressure time series as ground truth for prediction. PPGBP, instead, only includes two scalar values (SBP and DBP) per sample.
Thus, sig2sig models cannot be trained on this dataset.
All model performances are evaluated using the test set MAE on SBP and DBP separately.
The training protocol uses a 5-fold per-subject Cross-Validation for all datasets except UCI.
Given its size, single-held-out validation and test sets are adopted for UCI. 

Notably, cross-patient inference following these protocols yields significantly higher estimation errors than medical-grade device requirements, which can only be reached through personalized fine-tuning~\cite{almeidaAktiia2023}. However, this additional training is orthogonal to our work, which aims to demonstrate the feasibility of deploying efficient DNNs for BP estimation onboard wearable hardware.
\vspace{-0.1cm}
\begin{figure}[t]
    \centering
    \includegraphics[width=1.05\columnwidth]{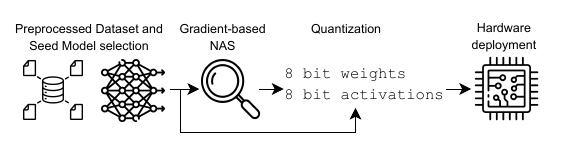}
    \caption{Overview of the proposed automated DNN optimization flow.}
    \label{fig:flow}
    \vspace{-0.6cm}
\end{figure}
\vspace{-0.1cm}
\subsection{Network Optimization}
To optimize our DNNs, we leverage the open-source library PLiNIO~\cite{plinio}, which provides an easy-to-use interface to implement various NAS and Quantization algorithms. A scheme of the optimization steps is shown in Fig.~\ref{fig:flow}.
Together with the training dataset, preprocessed and windowed as discussed above, the other key input of the pipeline is a \textit{seed} network, i.e., an initial DNN, which serves as a blueprint to generate optimized models. We use the two best-performing DNNs from~\cite{nature-survey} as seeds. Both are 1D CNNs, but while the first belongs to the scalar regressors category and is derived from a ResNet~\cite{resnet-ppg}, the second is a UNet-like~\cite{unet} sig2sig model.
We refer the reader to~\cite{nature-survey} for further details on the seeds.

Notably, these two architectures have already been optimized in~\cite{nature-survey} for each dataset, albeit only for maximizing accuracy. In contrast, in our work, we perform \textit{cost-aware} optimizations, showing that this permits us to find similarly or better performing models, which are additionally smaller and more efficient.

\subsubsection{NAS}
\begin{figure}
    \centering
    \includegraphics[width=.9\columnwidth]{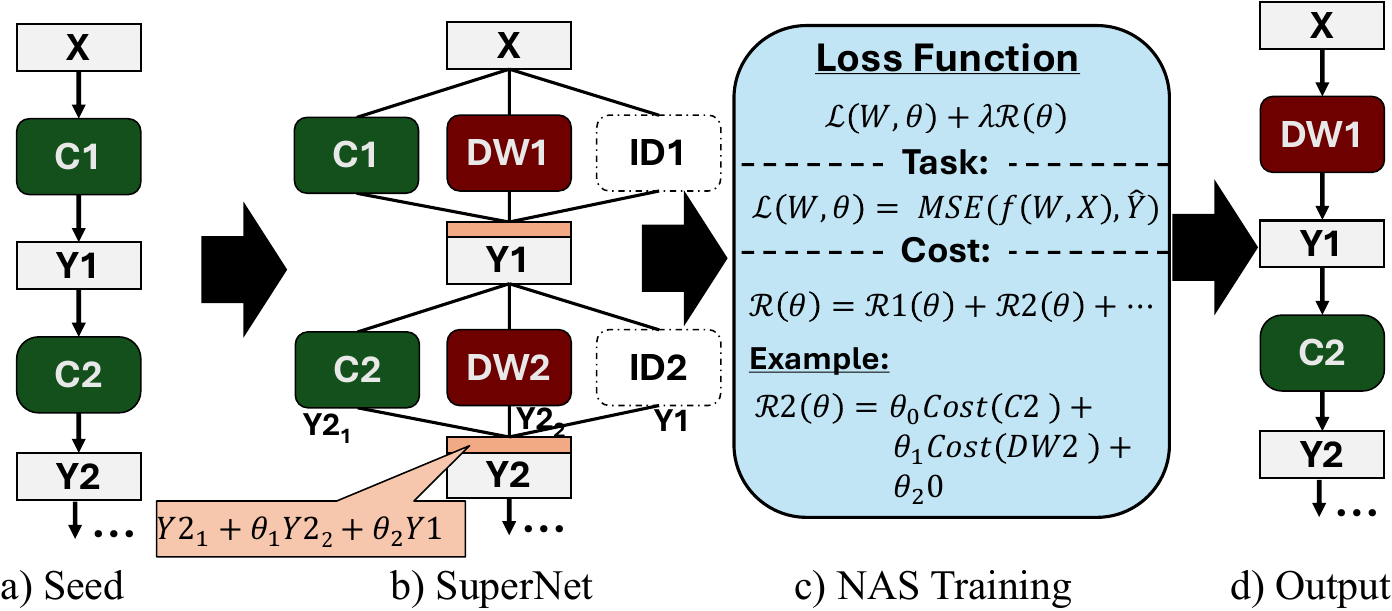}
    \vspace{-0.2cm}
    \caption{SuperNet-based NAS.}
    \label{fig:supernet}
    \vspace{-0.6cm}
\end{figure}

For each seed, the first optimization step consists of the application of a gradient-based NAS called \textit{SuperNet}, inspired from~\cite{darts}, whose working principle is depicted in Fig.~\ref{fig:supernet}. This method replaces each convolutional layer in the seeds with a pool of alternatives, all receiving the same input. The output of each layer is obtained as a linear combination of the various alternatives' outputs, weighted by softmax-ed trainable parameters $\theta_i$ (Fig.~\ref{fig:supernet}b).
Intuitively, finding a good architecture corresponds to setting, for each layer, one of the $\theta_i=1$ (and the others $=0$).
The NAS solves a continuous relaxation of this problem by inserting the multi-path DNN in a standard training loop, where both the normal network weights $W$ and the newly added $\theta$ are optimized jointly by gradient descent. 
This training uses the modified loss function shown in Fig.~\ref{fig:supernet}c, where $\mathcal{L}$ is the standard task loss, i.e., in our case, the Mean Squared Error (MSE) between the network's output $f(W,X)$ and the ground truth $\hat{Y}$. The newly added term $\mathcal{R}$, instead, is the \textit{expected cost} of the network as a function of the layer selection parameters. An example of its calculation is shown in Fig.~\ref{fig:supernet}c. In this work, we use \textit{model size} as a cost metric. At the end of the training, the output architecture is generated by selecting, for each layer, the alternative associated with the largest $\theta_i$. Varying the scalar \textit{regularization strength} $\lambda$, which controls the balance between the two loss terms, allows the generation of multiple output DNNs with different error vs cost trade-offs.

In our work, we use the SuperNet to select, for each layer, between a standard 1D convolution (C), a Depthwise-separable block (DW), and an identity operation (ID). The original models only include standard convolutions. The DW block, made of a sequence of a depthwise convolution and a pointwise layer, was first made popular by~\cite{mobilenet}, and has been shown to provide a lower-size yet similar-accuracy approximation of standard convolutions, leading to tiny yet capable networks. The ID, instead, is added only when input and output tensors have the same shape, and lets the NAS modulate the network depth by skipping some layers.

\subsubsection{Quantization}
In a second optimization step, we select some of the Pareto-optimal DNNs generated by the NAS and quantize them to \texttt{int8} format. For this, we use PLiNIO's Quantization-Aware Training (QAT) capabilities~\cite{jacobs2018}. We use a standard min-max affine quantization format for weights and the Parametrized Clipping Activation (PaCT) method for layer's inputs and outputs~\cite{pact}. Accumulation and biases are on 32 bits, as supported by our target inference library~\cite{pulp-nn}.

Note that the adopted NAS and quantization methods are not new per se. However, to our knowledge, we are the first to apply HW-aware optimizations for BP estimation.
\vspace{-0.1cm}
\subsection{Network Deployment}
We deploy our networks on the GreenWaves GAP8~\cite{gap8}, a low-power, RISC-V-based multi-core IoT processor designed specifically for signal processing tasks on edge devices. 
GAP8 features a cluster of eight general-purpose cores used to accelerate compute-intense workloads.
It also includes a 2-level scratchpad memory, with 512 kB of main memory, used to store the application code and DNN weights, and a 64 kB last-level cache with single-clock access latency for the cluster. A DMA engine moves the data between memory levels.

To convert our optimized DNNs into inference code for GAP8, we employ the DORY compiler~\cite{dory}. DORY automatically generates C code that handles the entire inference process, including memory management, DMA transfers scheduling, and optimized AI primitives invocation. It can directly take as input quantized DNNs generated by PLiNIO. As backend library for implementing each layer, we use~\cite{pulp-nn}.
We profile our deployed models on the GAP8 evaluation board, utilizing the internal performance counters for measuring latency, and the Nordic Power Profiler Kit II for power~\cite{nordic-2}.

\vspace{-0.3cm}
\section{Results}

\begin{figure*}
    \centering
    \includegraphics[width=.9\textwidth]{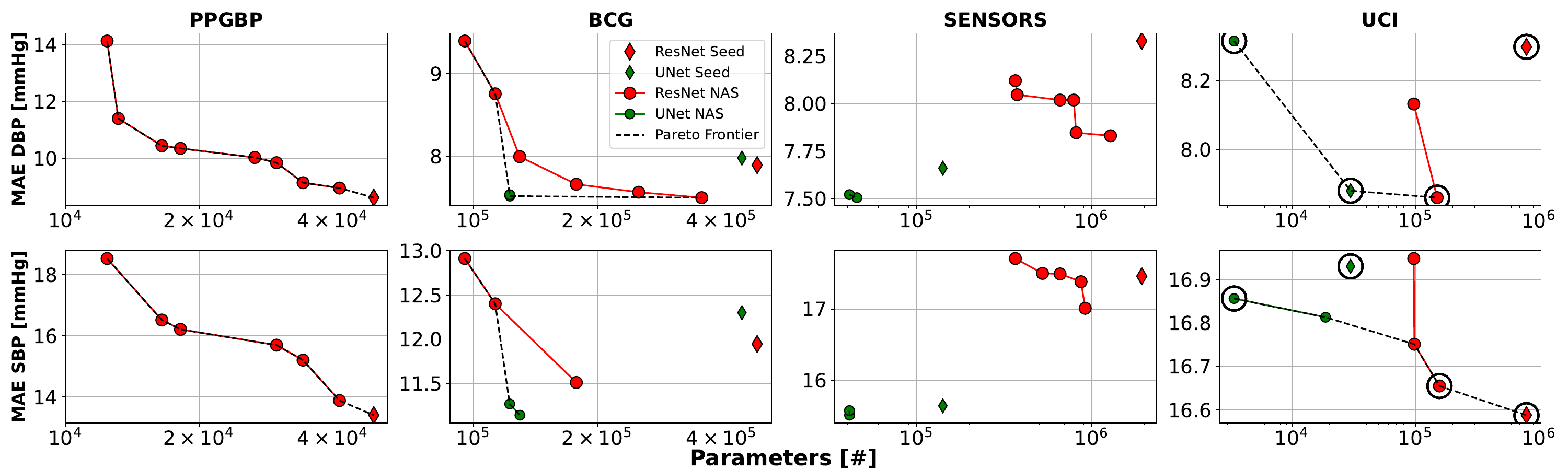}
    \vspace{-0.2cm}
    \caption{NAS results on all datasets and on DBP and SBP prediction.}
    \label{fig:res}
    \vspace{-0.6cm}
\end{figure*}
We performed all trainings using the Adam optimizer with a learning rate of 0.001 for the network weights and a separate Adam optimizer with a learning rate of 0.01 for the NAS parameters ($\theta$). In each epoch, the network weights were optimized on the training set, while the NAS parameters were optimized on the validation set, as in~\cite{darts}. For all datasets and both ResNet and UNet seeds, we tested 18 different values of $\lambda$, evenly spaced on a log scale between $10^{-11}$ and $10^{-7}$. 
We compared our optimized models with the original ResNet and UNet from~\cite{nature-survey}, as well as with two classic model families also considered in that paper, namely Random Forests (RFs) and Support Vector Regressors (SVR).
\vspace{-0.1cm}
\subsection{Pareto Analysis}
Fig.~\ref{fig:res} depicts the results of NAS on all four datasets and for DBP and SBP prediction. All results are reported in a MAE vs model size (n. of parameters) plane. Red and green diamonds correspond to the optimization seeds (ResNet and UNet from~\cite{nature-survey} respectively). Correspondingly colored circles are the Pareto-optimal architectures found with NAS. All results refer to floating point DNNs, before quantization.

On all datasets, we obtain models that either dominate the seeds or are on the memory vs error Pareto front. On the smallest one, PPGBP, we obtain a rich curve of Pareto architectures starting from the ResNet. 
We are able to reduce the seed size by 16\%, with a small increase in the MAE of only 3.9\% and 3.5\% on DBP and SBP prediction, respectively.
As mentioned, we do not have results with the UNet seed for PPGBP, given that this dataset does not include the full BP signal ground truth.
On BCG, we Pareto-dominate both seed networks, improving both MAE and size.
Our best UNet-derived model obtains 11.139 mmHg MAE on SBP prediction and 7.52 mmHg MAE on DBP, being 6.7\%/4.7\% better than the best seed (ResNet). Simultaneously, this network reduces the total number of parameters by 3.8$\times$.
However, it shall be noted that for these two datasets, classical ML methods still outperform our optimized DNNs in both performance and size, as reported in \cite{nature-survey}.
SVR achieves the lowest MAE in DBP estimation for both PPGBP and BCG datasets (8.04 and  7.34 mmHg, respectively) and a MAE of 13.15 mmHg and 11.45 mmHg on SBP estimation, being outperformed by the UNet, solely on BCG.

While being interesting for small datasets, classic ML models fail to benefit from the availability of larger amounts of data.
On the second largest dataset, Sensors, classical ML methods have slightly better performance than the seeds, but inferior to our NAS-optimized DNNs. Namely, SVR, which achieved the best results on both metrics (15.60 mmHg for SBP and 7.50 mmHg for DBP), is now outranked by our UNet NAS models (15.51 mmHg for SBP and the same DBP) with a parameters reduction up to 40$\times$.
On UCI, the dataset with the most samples, classic methods are outperformed even by the seeds, as shown in~\cite{nature-survey}; the best one (a RF) achieves a SBP MAE of 16.85 mmHg (versus 16.59 mmHg of the ResNet), while SVR is outperformed by UNet, with a DBP MAE of 8.07 vs 7.88 mmHg respectively. Moreover, the higher complexity of these datasets causes the number of parameters of both the SVR (with RBF kernel) and of the RF to increase exponentially. For instance, on UCI, the SVR becomes 998$\times$ larger than our best NAS output.

Conversely, on these two larger datasets, thanks to our NAS, we are again able to obtain Pareto-dominant solutions. On Sensors, our UNet-derived architectures reduce the size of the most accurate seed (UNet) by 3.4$\times$, while achieving a similar or lower MAE of 7.51 mmHg / 15.51 mmHg on DBP/SBP, respectively. 
Interestingly, on BCG and Sensors, Unet-based architectures outperform ResNets. We attribute this behavior to the ability of this network topology to learn faster from a lower amount of data, thanks to the richer training signal provided by the full time series reconstruction task.
The situation reverses in UCI, where ResNet-derived DNNs achieve the best performance.
The most accurate networks found with our NAS on UCI require only 149.8k/156.3k parameters to achieve a close-to-optimal MAE of 16.655 mmHg on SBP estimation, and the lowest overall (7.86 mmHg) on DBP estimation. 
While the seed ResNet is able to achieve an even lower MAE on SBP, with its 792k parameters, it would be impossible to deploy on GAP8's internal memory of 512KB, even when quantized.
\vspace{-0.1cm}
\subsection{Quantization \& Deployment}
\begin{table}[]
\caption{Deployment results on GAP8.}
\vspace{-0.15cm}
\label{tab:res-quant}
\footnotesize
\begin{tabular}{llllll}
\hline
\multicolumn{1}{l|}{Model}    & MAE-SBP       & \multicolumn{1}{l|}{MAE-DBP}       & Size [B] & Lat. [ms] & E. [mJ] \\ \hline
\multicolumn{6}{l}{Floating Point   Models (fp32)}                                                                         \\ \hline
\multicolumn{1}{l|}{ResNet}   & 16.59         & \multicolumn{1}{l|}{8.3}           & 3.17M    & n.a.         & n.a.        \\
\multicolumn{1}{l|}{UNet}    & 16.93         & \multicolumn{1}{l|}{7.88}          & 118.9k   & n.a.         & n.a.        \\ \hline
\multicolumn{6}{l}{Quantized Models   (int8)}                                                                              \\ \hline
\multicolumn{1}{l|}{ResNet}   & 18.23         & \multicolumn{1}{l|}{8.17}          & 791.8k   & o.o.m.        & o.o.m.        \\
\multicolumn{1}{l|}{UNet}    & 17.63         & \multicolumn{1}{l|}{8.19}          & 29.8k    & 7.04         & 0.36        \\
\multicolumn{1}{l|}{ResNet-B} & 17.83         & \multicolumn{1}{l|}{8.44}          & 156.3k   & 7.12         & 0.36        \\
\multicolumn{1}{l|}{Resnet-S} & 17.48         & \multicolumn{1}{l|}{\textbf{8.08}} & 149.8k   & 7.27         & 0.37        \\
\multicolumn{1}{l|}{UNet-S}  & \textbf{17.2} & \multicolumn{1}{l|}{8.26}          & 23.4k    & 8.91         & 0.45        \\ \hline
\end{tabular}
\vspace{-0.6cm}
\end{table}

For the sake of space, we report our deployment results only on the largest and most challenging dataset, UCI. We quantize and deploy the DNNs marked with black circles in Fig.~\ref{fig:res}. Namely, the two seeds, and the NAS outputs at the extremes of the Pareto front.
All the results are reported in Table \ref{tab:res-quant}, where \emph{ResNet-B} and \emph{ResNet-S} are the biggest NAS models on the Pareto front of the SBP and DBP graphs, respectively; \emph{UNet-S} is the smallest Pareto model, which has the same architecture for both SBP and DBP estimation.
The table reports the error, size, latency (Lat.), and energy per inference (E.) for each model. For reference, the MAE and size of the seeds in floating point are also reported, although these models are not deployable on the FPU-less GAP8.
The size and latency reductions, thanks to quantization, are paid with a slight increase in MAE (up to 9.8\%).
ResNet models tend to be more susceptible to this degradation. 
After quantization and deployment, the best results are achieved by the Resnet-S, which achieves an 8.08 mmHg MAE on DBP estimation, and by the UNet-S, achieving 17.2 mmHg of MAE on SBP estimation.
Due to the too-high number of parameters, the seed ResNet can not be deployed on GAP8's onboard memory. On the other hand, all NAS produced models fit the platform.
Compared to the seed UNet, we achieve a similar latency and energy consumption; the UNet-S model, which achieves a latency of 8.91 ms with an energy consumption of as low as 0.45 mJ, is indeed made mostly of DW layers, which are smaller but also less efficient when deployed, reducing the memory occupation at the cost of a limited increase in latency.

\vspace{-0.2cm}
\section{Conclusion}
\vspace{-0.1cm}
The efficient execution of PPG-based BP estimation algorithms is critical for the prevention of important diseases associated, for instance, to hypertension.
With our experiments, we demonstrated the possibility of embedding accurate DNN models on low-power wearable-class devices, achieving SoA performance.
Future work will focus on the fine-tuning of our models on patient-specific data to reach competitive accuracy with the golden standard of non-intrusive PB.

\bibliographystyle{IEEEtran} %
\bibliography{bstctl,bibliography} %
\vspace{12pt}

\end{document}